\documentclass[12pt,preprint]{aastex}

\usepackage{epsfig}

\shorttitle{He I Case B Emissivities}
\shortauthors{Porter et al.}

\begin{document}
\title{Theoretical He I Emissivities in the Case B Approximation}


\author{R. L. Porter, R. P. Bauman, G. J. Ferland, \& K. B. MacAdam}
\affil{Dept. of Physics and Astronomy, University of Kentucky, Lexington, KY, 40506}
\email{rporter@pa.uky.edu}

\begin{abstract}
We calculate the He I case B recombination cascade spectrum using improved radiative and
collisional data.  We present new emissivities over a range of electron temperatures and densities.
The differences between our results and the current standard are large enough to have a significant
effect not only on the interpretation of observed spectra of a wide variety of objects but also on
determinations of the primordial helium abundance.
\end{abstract}


\section{Introduction}

Helium is the second most abundant element in the universe, and its emission
and opacity help determine the structure of any interstellar cloud. Its abundance
relative to hydrogen can be measured within a few percent since the emissivities
of H I and He I lines have similar dependences on temperature and density.
This makes it an indicator of both stellar and primordial nucleosynthesis (Pagel 1997).

A good discussion of the history of calculations of the helium recombination
spectra is given by Benjamin, Skillman, {\&} Smits (1999, hereafter BSS99),
who present new calculations - the current standard in the field. Yet much
progress has been made since the work by Smits (1991, 1996) upon which the BSS99
results depend. We implement these improvements, present a new set of predictions, 
and compare our results with those of BSS99.  The differences are large enough to
impact continuing attempts to estimate the primordial helium abundance
(Peimbert \textit{et al} 2002).

\section{The New Model Helium Atom}

The basic physical processes have been described by Brocklehurst (1972) and BSS99.
Here we will describe the differences between BSS99 and our new numerical representation of
the helium atom, which is a part of the spectral simulation code Cloudy (Ferland \textit{et al} 1998).
This model resolves all terms, $nlS$, up to an adjustable maximum principal
quantum number $n_{max}$, followed by a pseudolevel, $n_{max}+1$,
in which all $lS$ terms are assumed to be populated according to statistical
weight and ``collapsed" into one. We set recombinations into the collapsed level
equal to the convergent sum of recombinations from $n=n_{max}+1$ to $\infty$.  In the
low-density limit, the collapsed level increases the emissivities of our
benchmark lines (the same 32 lines given in BSS99) by 0.4\%, on average, with
$n_{max}$=100.  The decays from states with $l=n-1$ are most sensitive to this correction
for system truncation. The strong optical line $\lambda5876$ is corrected upward by 1.3\%.
At finite densities collisional processes force the populations of very highly excited
states into local thermodynamic equilibrium (LTE).  In this case the adequacy of
the method used to compensate for truncation is unimportant.  We find the corrections
negligible for $n_{e}=100$ cm$^{-3}$ and $n_{max}$=100.  Consequently, the uncertainties
in the results presented in Section 3 are due to the uncertainties in atomic data,
especially the often substantial uncertainties in collisional rates affecting
terms not in LTE at given conditions. 
  
There are several differences in atomic data for radiative processes between BSS99 and
the present work.  The transition probabilities and radiative recombination coefficients are
obtained from oscillator strengths and photoionization cross-sections.
BSS99 uses the oscillator strengths calculated by Kono {\&} Hattori (1984).
While these agree very well with the essentially exact
oscillator strengths of Drake (1996), Drake presents a much larger set, up to and
including $n$=10 and $l$=7, which we adopt. Hummer {\&} Storey (1998, hereafter HS98)
have presented \textit{ab initio} calculations of threshold photoionization
cross-sections up to $n$=4.  BSS99 uses cross-sections from
TOPbase\footnote[1]{http://vizier.u-strasbg.fr/topbase/topbase.html}
(Cunto, 1993), while we use the more accurate HS98 values.
The dominant remaining uncertainties in radiative data are in
oscillator strengths involving low $l$ states (with $n>10$) and photoionization cross-sections for low $l$
states (with $n>4$). HS98 also illustrates the method, originally discussed by
Seaton (1958), of calculating threshold photoionization cross-sections by
extrapolating absorption oscillator strengths to the threshold energy of a given level.
This method has been used in the present work, based on the oscillator strengths from Drake,
to extend the \textit{ab initio} cross-sections of HS98 to greater $n$.

Differences in collisional data between BSS99 and the current work are also significant.
For low-\textit{n} transitions for which there are \textit{ab initio} calculations,
BSS99 uses the collision strengths of Sawey {\&} Berrington (1993). We replace these,
where available, with the results of the close-coupling calculation by
Bray \textit{et al} (2000), which include continuum states not considered in the
$R$-matrix calculations by Sawey {\&} Berrington.
For \textit{l}-changing collisions BSS99 uses two different treatments: Seaton (1962, hereafter S62)
for low-$l$ transitions, and Pengelly {\&} Seaton (1964, hereafter PS64) otherwise.  Neither of these
treatments allows for angular momentum transfers greater than one unit, and both apply when the 
projectile velocity is greater than the velocity of the bound electron.
The r.m.s. electron and projectile velocities in conditions considered by BSS99,
assuming proton colliders, are

\begin{equation}
\label{v_e} 
v_e = \frac{Z\alpha c}{n} \mbox{  and  } v_{proj} = \sqrt {\frac{3kT}{m_{p}}},
\end{equation}

\noindent
where $\alpha$ is the fine structure constant, \textit{c} is the speed of light, and $Z$ is
the screened nuclear charge. Dividing the latter by the former, we arrive at the expected value of the 
reduced velocity as a function of temperature and principal quantum number

\begin{equation}
\label{vtilde} 
 \langle\tilde {v}\rangle = \frac{v_{proj} }{v_e } = 7.19\times10^{ - 5}n\sqrt{T({\rm K})}.
\end{equation}

\noindent
For typical nebular temperatures this reduced velocity will be of order or 
less than unity for proton colliders for all $n  \le $ 150; the treatment of PS64 is applicable
only for greater $n$, and it greatly overestimates \textit{l}-mixing cross-sections
when used outside its range of validity (MacAdam, Rolfes, \& Crosby 1981).

\begin{figure*}
	\includegraphics*[angle=-90,width=6.5in,keepaspectratio=true]{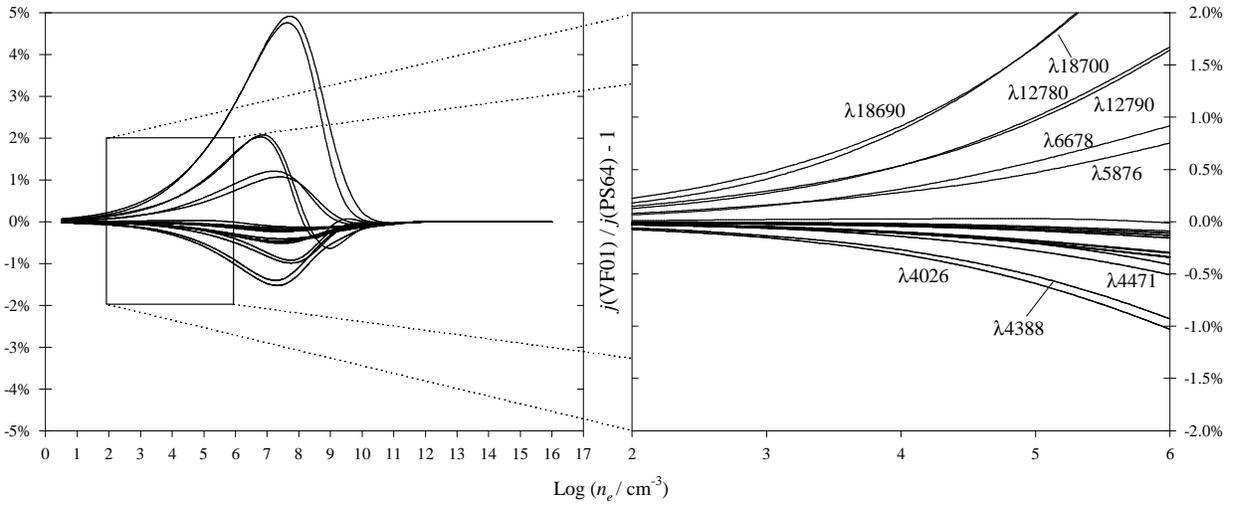}
\caption{\footnotesize 
Percent difference between the emissivities calculated using two different
Stark collision treatments, for several strong lines, as a function of $n_{e}$.
Left panel: a wide range of densities - as expected, there is no effect in either
the low density extreme, because the collision rates are negligible, or the high
density extreme, where the Stark collisions force the terms to LTE. The majority
of lines are most sensitive at densities found in stellar envelopes and quasar
emission line regions. Right panel: the range of densities found in nebulae -
several lines have a sensitivity to the Stark collision treatment of about 1{\%}.}
\label{vf_vs_ps}
\end{figure*}

Vrinceanu {\&} Flannery (2001, hereafter VF01) give a (classical) theory of \textit{l}-changing
collisions and claim exact solutions in the limit that the intrashell transition
is induced by slow distant collisions. Their treatment allows naturally for angular
momentum changes greater than unity. (At a sufficiently high reduced velocity large
angular momentum transfers are strongly suppressed and the theory goes to the optically
allowed limit with which PS64 is concerned.) We use equation 41 of Kazansky {\&} Ostrovsky (1996)
for the angle, $\Delta\Phi$, swept out by the projectile.  A physical basis for the necessary
large impact parameter cutoff in the theory follows from equating the Stark and quantum-defect
precession frequencies.  The Stark frequency is given by

\begin{equation}
\label{omega_s} 
\omega _s = \frac{3Z_1 n}{2b^2} a.u.,
\end{equation}

\noindent
and the quantum-defect precession frequency
(Hezel \textit{et al} 1992) is given by

\begin{equation}
\label{omega_qd} 
\omega _{qd} = \frac{5\delta _l}{n^3l}\left(1-\frac{3l^2}{5n^2}\right) a.u.
\end{equation}

\noindent
where \textit{$\delta $}$_{l}$ is the quantum defect, $Z_{1}$ is the charge of the projectile
and $b$ is the impact parameter.  
By setting $\omega _{qd}$ equal to $\omega _s$, we obtain a maximum impact parameter, $b_{max}$.
The electron orbit precession will be faster than the Stark beating at larger impact parameters,
so that transitions are increasingly less likely.  To insure symmetry, we use the average 
$\omega _{qd}$ of the initial and final levels.  
We use VF01 for $l$-changing collisions involving initial and final levels with $l\ge 3$,
and like BSS99 we use the impact parameter treatment of S62 for $l$-change from $s$, $p$, and $d$ levels.
We use electron, proton and He$^+$ colliders for all transitions, taking
$n_{He^+}$ = 0.1 $n_{p}$ and $n_{e}$ = $n_{p}$ + $n_{He^+}$. Since S62, which describes electron collisions,
is based upon the method of virtual quanta (see Jackson 1999), we can readily adapt it
for the positive-ion collisions: The power spectrum of the time-dependent fields generated
at the target atom by a passing charged projectile depends only on the projectile's charge magnitude,
speed (not kinetic energy or mass separately) and impact parameter. The same considerations apply to
PS64 and VF01 and have been implemented to allow for all three collider species. In calculating
the necessary thermal averages we have assumed that the same temperature characterizes electrons,
protons and He$^+$ ions.

Figure \ref{vf_vs_ps} compares emissivities we predicted using the VF01 and PS64 theories.
The predicted emissivities typically change by about 1{\%} for nebular densities
by using the theory of VF01 rather than that of PS64. The difference is much
greater at high densities found, for example, in parts of quasars.

\section{Results}

\begin{figure}[t]
	\includegraphics[angle=-90,width=6.5in,keepaspectratio=true]{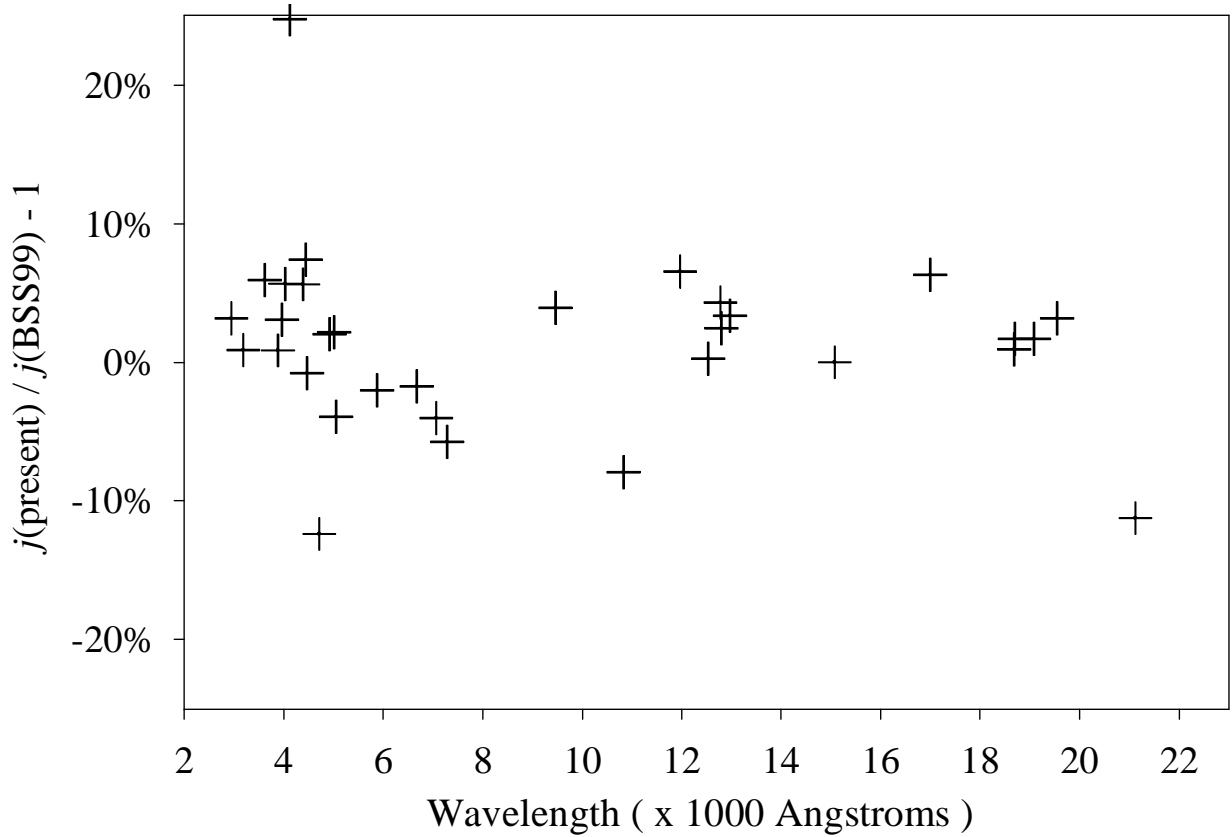}
\caption{\footnotesize A comparison of the present results with those of BSS99 at $T_{e}=10^{4}$ K
and $n_{e}=10^{4}$ cm$^{ - 3}$.}
\label{comp_w_bss}
\end{figure}


In Figure \ref{comp_w_bss} we compare our results with those of BSS99 for the case
$T_{e}=10^{4}$ K and $n_{e}=10^{4}$ cm$^{ - 3}$. The average difference for the 32 emission lines is 4.6{\%}.
The greatest difference is for $\lambda$4121, for which our emissivity is 25{\%} greater.
In general, agreement worsens with increasing density; at $n_{e}=10^{2}$ cm$^{ - 3}$, 
the average and greatest differences are 1.6{\%} and 6.4{\%}, respectively,
while at $n_{e}=10^{6}$ cm$^{ - 3}$, we find differences of 7.0{\%} and 35{\%}.  Agreement also
worsens with increasing temperature.  Table \ref{tab1} presents emissivities for all
of the temperatures and densities considered by BSS99.  We believe that these results
are a significant improvement. The application of these results to specific astrophysical
problems will be the subject of future papers.

We thank G. W. F. Drake for making available extensive tables of his calculations, D. Vrinceanu
and M. J. Cavagnero for helpful discussions, and P. J. Storey, whose constructive criticisms
helped us to significantly improve our work. We also acknowledge support from NASA grant NAG5-12020 and
NSF grant AST 0307720.

\begin{deluxetable}{rlrrrrrrrrr}
\tabletypesize{\scriptsize}
\tablecaption{He I Case B Emissivities.\label{tab1} }
\tablewidth{0pt}
\tablehead{
	\colhead{}   					&
	\colhead{$T_e\ ({\rm K})$: }			&
	\multicolumn{3}{l}{5000}			&
	\multicolumn{3}{l}{10000}			&
	\multicolumn{3}{l}{20000}			\\
	\colhead{$\lambda $({\AA})} 			& 
	\colhead{$n_e\ ({\rm cm}^{-3})$:}		& 
	\colhead{10$^{2}$}				& 
	\colhead{10$^{4}$}				& 
	\colhead{10$^{6}$}				& 
	\colhead{10$^{2}$}				& 
	\colhead{10$^{4}$}				& 
	\colhead{10$^{6}$}				& 
	\colhead{10$^{2}$}				& 
	\colhead{10$^{4}$}				& 
	\colhead{10$^{6}$}
}
\startdata
2945& & 0.4142& 0.4261& 0.4567& 0.2687& 0.2816& 0.2958& 0.1648& 0.1987& 0.2112\\
3187& & 0.8693& 0.8950& 0.9594& 0.5617& 0.6119& 0.6507& 0.3432& 0.4584& 0.4963\\
3614& & 0.1115& 0.1151& 0.1241& 0.0691& 0.0717& 0.0752& 0.0397& 0.0471& 0.0496\\
3889& & 2.2452& 2.3261& 2.5038& 1.4116& 1.6794& 1.8348& 0.8315& 1.3332& 1.4859\\
3965& & 0.2280& 0.2353& 0.2532& 0.1409& 0.1471& 0.1543& 0.0807& 0.0966& 0.1020\\
4026& & 0.5279& 0.5427& 0.5866& 0.2917& 0.3029& 0.3175& 0.1457& 0.1782& 0.1896\\
4121& & 0.0341& 0.0348& 0.0363& 0.0249& 0.0300& 0.0323& 0.0184& 0.0338& 0.0379\\
4388& & 0.1411& 0.1453& 0.1568& 0.0772& 0.0798& 0.0834& 0.0380& 0.0445& 0.0468\\
4438& & 0.0145& 0.0149& 0.0157& 0.0101& 0.0111& 0.0117& 0.0070& 0.0100& 0.0107\\
4471& & 1.1469& 1.1781& 1.2699& 0.6124& 0.6465& 0.6806& 0.3010& 0.4077& 0.4418\\
4713& & 0.0904& 0.0929& 0.0977& 0.0652& 0.0833& 0.0917& 0.0478& 0.0957& 0.1087\\
4922& & 0.3132& 0.3222& 0.3463& 0.1655& 0.1722& 0.1801& 0.0799& 0.0973& 0.1027\\
5016& & 0.5849& 0.6039& 0.6498& 0.3539& 0.3808& 0.4035& 0.1996& 0.2554& 0.2735\\
5048& & 0.0355& 0.0366& 0.0387& 0.0244& 0.0281& 0.0301& 0.0167& 0.0253& 0.0276\\
5876& & 3.3613& 3.4419& 3.6889& 1.6344& 1.8724& 2.0179& 0.7887& 1.4753& 1.6649\\
6678& & 0.9640& 0.9872& 1.0512& 0.4629& 0.4962& 0.5223& 0.2170& 0.3082& 0.3250\\
7065& & 0.4273& 0.4750& 0.5303& 0.2997& 0.5897& 0.7166& 0.2154& 0.6809& 0.8096\\
7281& & 0.1318& 0.1387& 0.1497& 0.0886& 0.1203& 0.1357& 0.0593& 0.1086& 0.1227\\
9464& & 0.0229& 0.0235& 0.0252& 0.0148& 0.0155& 0.0163& 0.0091& 0.0109& 0.0116\\
10830& & 4.9896& 13.9601& 21.4091& 3.3152& 18.8390& 25.4507& 2.3419& 18.8421& 23.6646\\
11970& & 0.0532& 0.0547& 0.0591& 0.0294& 0.0305& 0.0320& 0.0146& 0.0179& 0.0191\\
12530& & 0.0278& 0.0286& 0.0307& 0.0179& 0.0195& 0.0208& 0.0109& 0.0146& 0.0158\\
12780& & 0.2010& 0.2056& 0.2196& 0.0936& 0.0972& 0.1011& 0.0410& 0.0513& 0.0546\\
12800& & 0.0670& 0.0686& 0.0728& 0.0311& 0.0320& 0.0330& 0.0137& 0.0169& 0.0175\\
12970& & 0.0178& 0.0183& 0.0198& 0.0097& 0.0101& 0.0105& 0.0048& 0.0056& 0.0059\\
15080& & 0.0121& 0.0125& 0.0134& 0.0074& 0.0078& 0.0082& 0.0042& 0.0051& 0.0054\\
17000& & 0.0811& 0.0833& 0.0898& 0.0433& 0.0457& 0.0481& 0.0212& 0.0288& 0.0312\\
18680& & 0.5062& 0.5126& 0.5412& 0.2184& 0.2282& 0.2367& 0.0955& 0.1261& 0.1324\\
18700& & 0.1687& 0.1711& 0.1793& 0.0727& 0.0748& 0.0768& 0.0321& 0.0449& 0.0445\\
19090& & 0.0289& 0.0297& 0.0319& 0.0152& 0.0159& 0.0166& 0.0073& 0.0089& 0.0094\\
19550& & 0.0162& 0.0167& 0.0179& 0.0104& 0.0114& 0.0121& 0.0064& 0.0085& 0.0092\\
21120& & 0.0138& 0.0141& 0.0149& 0.0099& 0.0127& 0.0139& 0.0073& 0.0146& 0.0166\\
\enddata

\tablecomments{ Emissivities $4\pi j_{\lambda}/n_{e}n_{He^+}$ are given in units $10^{-25}$ erg cm$^{3}$ sec$^{-1}$.
The hydrogen density is 0.9 $n_{e}$, and the helium abundance is one-tenth of the hydrogen abundance.}


\end{deluxetable}

\end{document}